# A Scalable High-Performance Priority Encoder Using 1D-Array to 2D-Array Conversion

Xuan-Thuan Nguyen, *Student Member, IEEE*, Hong-Thu Nguyen, and Cong-Kha Pham, *Member, IEEE*

***Abstract*—In our prior study of an *L*-bit priority encoder (PE), a so-called one-directional-array to two-directional-array conversion method is deployed to turn an *L*-bit input data into an $M \times N$-bit matrix. Following this, an *N*-bit PE and an *M*-bit PE are employed to obtain a row index and column index. From those, the highest priority bit of *L*-bit input data is achieved. This brief extends our previous work to construct a scalable architecture of high-performance large-sized PEs. An optimum pair of $(M, N)$ and look-ahead signal are proposed to improve the overall PE performance significantly. The evaluation is achieved by implementing a variety of PEs whose *L* varies from 4-bit to 4096-bit in 180-nm CMOS technology. According to post-place-and-route simulation results, at PE size of 64 bits, 256 bits, and 2048 bits the operating frequencies reach 649 MHz, 520 MHz, and 370 MHz, which are 1.2 times, 1.5 times, and 1.4 times, as high as state-of-the-art ones.**

## I. Introduction

PRIORITY encoder (PE) is a particular circuit that resolves the highest priority match and outputs a matching location, or address, into binary format, from which corresponding data can be retrieved correctly. High-performance PEs have become increasingly important, especially for processing a massive amount of data in real time. Although some improvements in conventional PE are properly applied in advanced circuits, such as incrementer/decrementer [1], comparator [2], and ternary content-addressable memory [3], [4], the performance of those PEs deteriorates rapidly as their input sizes increase by several hundred bits.

Several hierarchical architectures have been proposed to manage large-sized PEs whose sizes reach to several thousand bits. An approach adopting a set of one-hot encoders [7] or a set of specific comparator and sort circuits [8] are the cases in point. Nonetheless, those architectures require many resources to maintain a sufficient operating frequency (FREQ).

Manuscript received October 5, 2016; revised January 13, 2017 and February 6, 2017; accepted February 19, 2017. Date of publication February 22, 2017; date of current version August 25, 2017. This work was supported in part by the VLSI Design and Education Center, in part by the University of Tokyo in collaboration with Synopsys, Inc., and in part by the Cadence Design Systems, Inc. This brief was recommended by Associate Editor A. J. Acosta.

The authors are with the Department of Engineering Science, University of Electro-Communications, Tokyo 182-8585, Japan (e-mail: xuanthuan@vlsilab.ee.uec.ac.jp).

Therefore, in this brief, we propose a set of principles, extending our previous 1D-to-2D conversion based PE [9], to construct a scalable high-performance PE. Our contribution focuses on:

- A methodology to build a 4-bit PE, an 8-bit PE, and a 16-bit PE, from which a large-sized PE will be created.
- A methodology to select optimum values of *M* and *N* for high-performance achievement.
- A methodology to reduce overall latency by using a look-ahead signal with an alternative multiplexer.

The proposed PEs are implemented in 180-nm CMOS process at different sizes, i.e., from 4-bit to 4,096-bit. Both *M* and *N* are also adjusted to observe the variation in PE performance. According to post-place-and-route simulation results, any PE deploying a 4-bit PE to generate a column index ($M = 4$) presumably attains the highest FREQ. Additionally, 1D-to-2D conversion significantly improves the deterioration in FREQ when rising PE size. In comparison with the state-of-the-art, the FREQs of our 64-bit PE, 256-bit PE, and 2,048-bit PE exceed 1.2 times, 1.5 times, and 1.4 times, respectively.

The remainder of this brief is organized as follows. Section II briefly summarizes previous approaches. Section III clearly describes a hardware architecture of large-sized PEs. Section IV shows the reported FREQ and resource in comparison with other designs. Lastly, Section V presents our conclusion.

## II. Previous Works

Fig. 1(a) illustrates a conventional architecture of PE64, including a set of prioritizers (PRIs) and encoder (ENC). $PRI_{i+1}$ is enabled by a control signal *C* from $PRI_i$, and so forth. Initially, 64-bit input data is split into eight 8-bit groups. Each PRI resolves the highest priority bit of each group, while ENC outputs a matching location into binary format. For instance, if $D_0$ is 01001110, $EP_0$ and *Q* become 0100000 and 000010, respectively. Because all PRI modules are connected in series, the worst latency of PE64 is about eight times as high as that of one PRI.

To reduce such latency, Huang *et al.* [1] presented multi-level lookahead and multi-level folding techniques. By remapping all control signals, the performance was improved up to ten times. However, this mapping strategy became increasingly complicated as PE size went up. Fig. 1(b) depicts a parallel priority look-ahead architecture, which was initially introduced by Kun *et al.* [3] and then was applied in ternary content-addressable memory [4]. With this architecture, $PRI_0$ to $PRI_7$ can return their priority matches in parallel due to the control signal provided by $PRI_8$. Despite decreasing the latency, the resource utilization rises because of the additional $PRI_8$ and logic gates. Another improvement from Balobas and Konofaos [6] exploited a new design of 4-bit PE (PE4) and a static-dynamic parallel priority lookahead architecture to boost the performance of PE64. However, the architectures of large-sized PEs were not mentioned. Furthermore, Abdel-Hafeez and Harb [5] presented a special prefix scheme for PEs whose size rises to 256 bits. Nevertheless, the performance declines sharply with increased PE size.

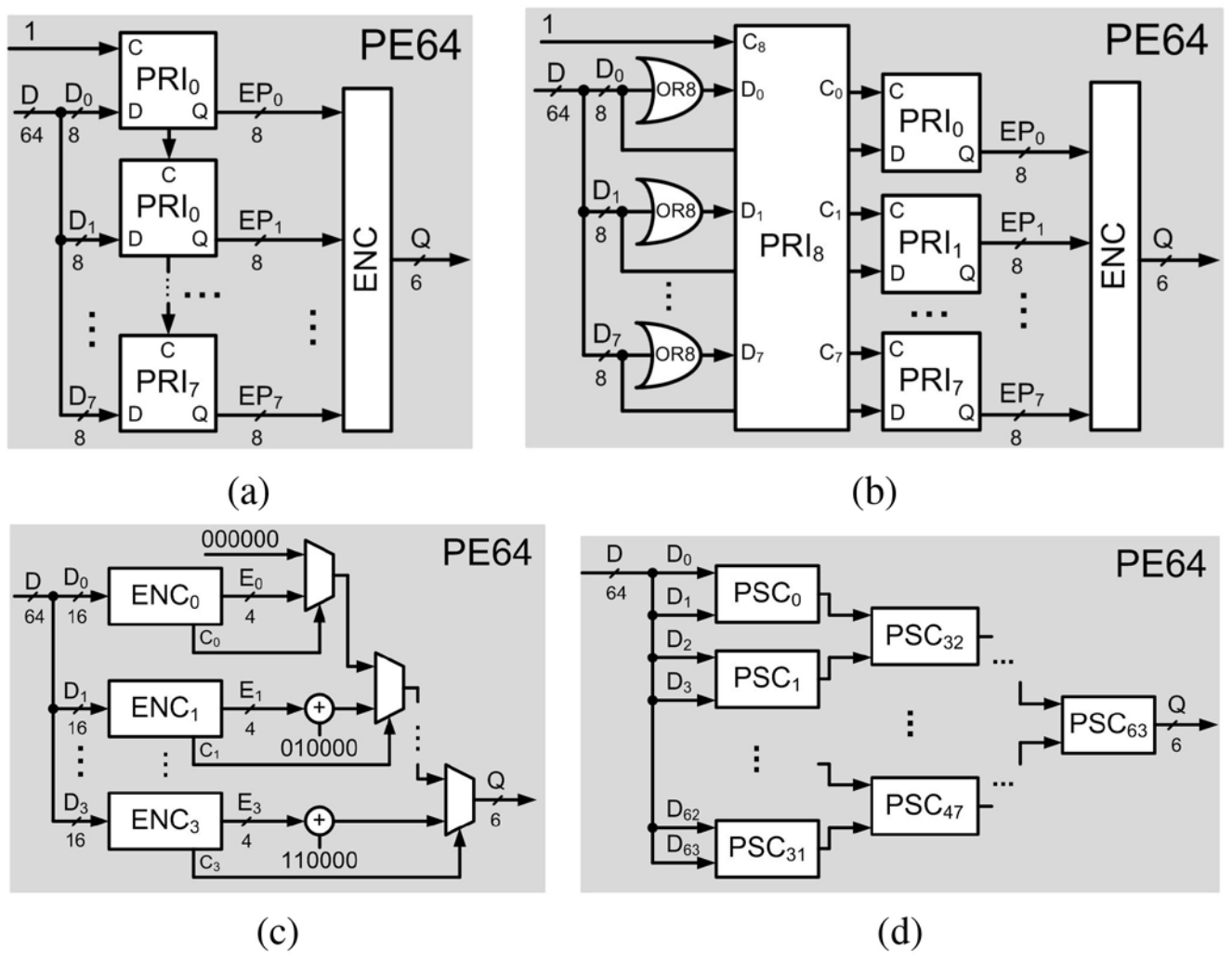

Fig. 1. The architecture of (a) conventional PE64, (b) parallel PE64, (c) PE64-based one-hot encoder, and (d) PE64-based comparison and sort circuit.

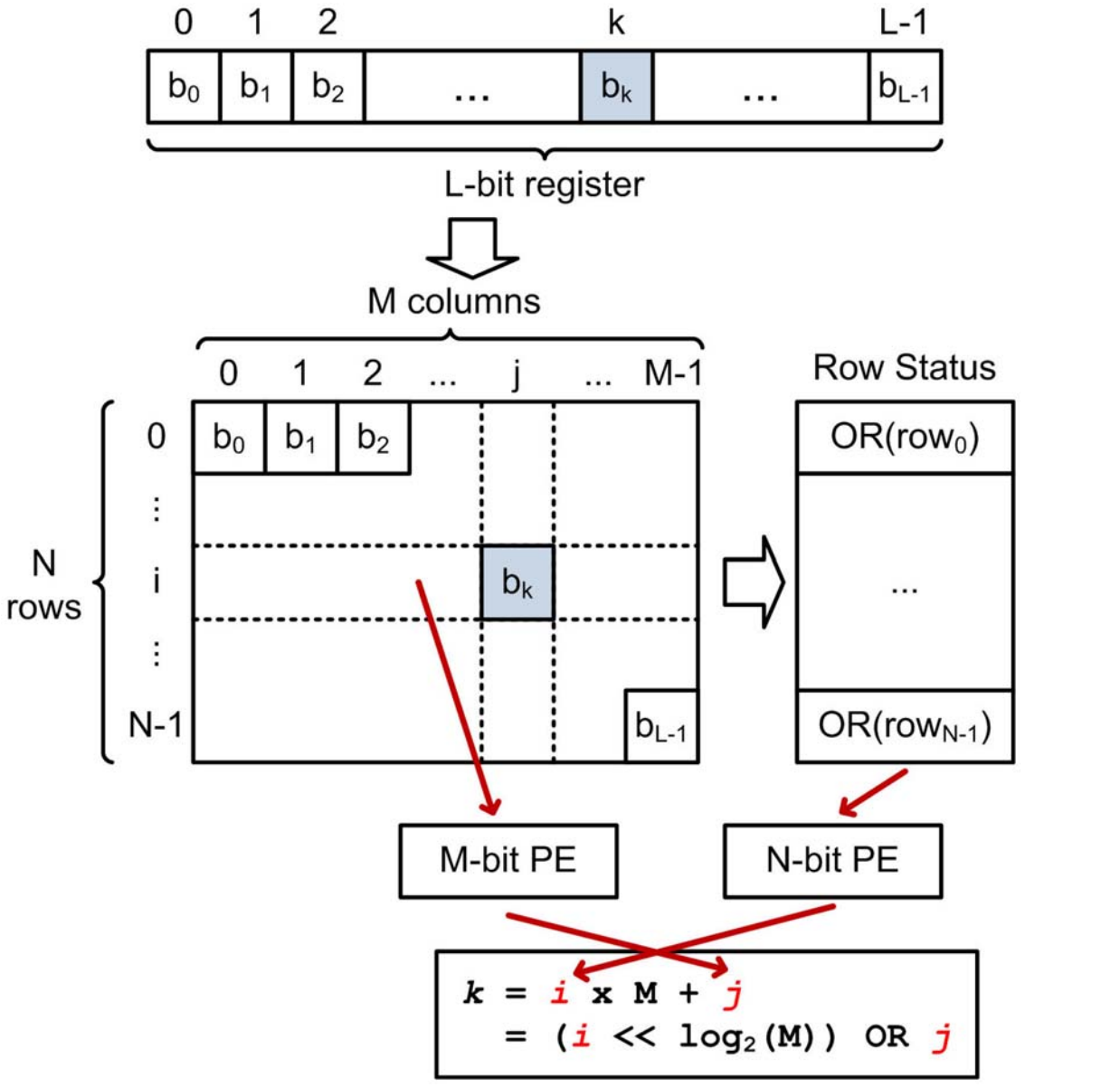

Fig. 2. The conversion from $L$-bit input to $M \times N$-bit input.

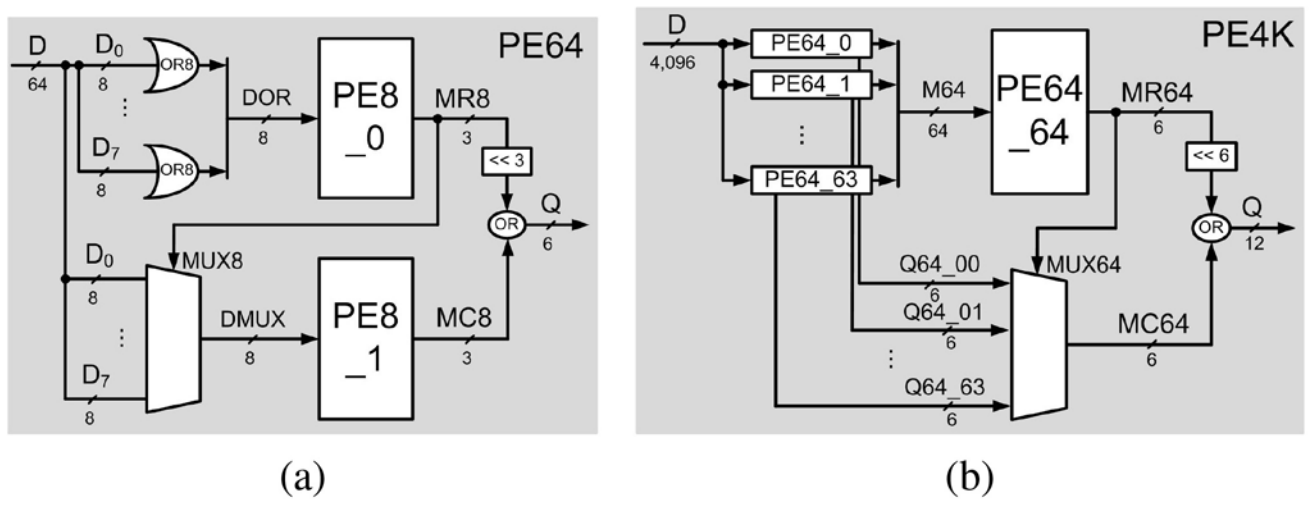

Fig. 3. The architecture of 1D-to-2D conversion based (a) PE64 and (b) PE4K.

Fig. 1(c) shows the architecture of a PE64 based on four one-hot encoders, which was designed by Le *et al.* [7]. Each ENC converts a corresponding 16-bit group into 4-bit position and a control signal $C$ decides whether the results are passed to next multiplexers. Suppose that PE size is 2,048 bits, up to 128 ENCs connected in series would be required. Fig. 1(d) depicts another approach proposed by Maurya and Clark [8], where a set of comparator and sort circuits (PSC) are deployed to check each pair of bits of input data so the highest priority bit is decided. If the PE size is 2,048 bits, as many as 2,047 PSCs connecting in 11 pipeline stages are demanded. In other words, those architectures are confronted for large-scale resource consumption.

A novel architecture of an $L$-bit PE using the 1D-to-2D conversion method was originally proposed in our previous work [9]. Fig. 2 illustrates this method, where $L$-bit input data is converted into a $M \times N$-bit matrix, with $M$ and $N$ are the numbers of columns and rows, respectively. All bits of row status are obtained by performing the bitwise OR to all bits in the corresponding row. Subsequently, an $N$-bit PE finds the highest priority bit $i$ (row index) in the $N$-bit row status, and an $M$-bit PE seeks the highest priority bit $j$ (column index) in this row $i$. The matching position $k$ of an 1D-array input is retrieved as $k = i \times M + j$. More significantly, if $M$ is a power of two, the multiplier and adder are simply replaced by the fixed wirings that function as left-shift and OR operators.

An architecture of 1D-to-2D conversion based PE64 is shown in Fig. 3(a), where 64-bit input data is considered as an $8 \times 8$-bit array. Two PE8s were then used to calculate indexes of row and column. From those, a location of highest priority bit was obtained. Similarly, a large-sized PE such as 4,096-bit PE (PE4K) was built by 64 PE64s connecting in parallel and one central PE64, as depicted in Fig. 3(b). Experimental results on multi-match priority encoders proved that at the size of 64-bit and 2,048-bit, our FREQs surpass those of [4] (1.7 times) and [7] (1.4 times), respectively. Nonetheless, an optimized architecture for high FREQ is still undiscovered. As a result, Section III will present a systematic approach to the scalable high-performance large-sized PEs in detail.

## III. Implementation

### A. Overview

Taking the example above, at PE size $L$ of 64-bit, ($M$, $N$) includes such values as (2, 32), (32, 2), (8, 8), (4, 16), and (16, 4). Selecting an optimum pair of ($M$, $N$) therefore plays an important role in constructing high-performance PEs.

### B. Architecture

Fig. 4(a) depicts the truth table and Boolean expression of a PE4. Similarly, the expressions of a PE8 and 16-bit PE (PE16) are correspondingly given in Fig. 4(b) and Fig. 4(c). We can

TRUTH TABLE

| $D_3$ | $D_2$ | $D_1$ | $D_0$ | $Q_1$ | $Q_0$ |
|---|---|---|---|---|---|
| 0 | 0 | 0 | 1 | 0 | 0 |
| 0 | 0 | 1 | x | 0 | 1 |
| 0 | 1 | x | x | 1 | 0 |
| 1 | x | x | x | 1 | 1 |

x: Don't care

$$Q_0 = \sum(D_1, D_3) = \overline{D_2}.D_1 + D_3$$
$$Q_1 = \sum(D_2, D_3) = D_2 + D_3$$

(a)

$$Q_0 = \sum(D_1, D_3, D_5, D_7) = \overline{D_6}.(\overline{D_4}.\overline{D_2}.D_1 + \overline{D_4}.D_3 + D_5) + D_7$$
$$Q_1 = \sum(D_2, D_3, D_6, D_7) = \overline{D_5}.\overline{D_4}.(D_2 + D_3) + D_6 + D_7$$
$$Q_2 = \sum(D_4, D_5, D_6, D_7) = D_4 + D_5 + D_6 + D_7$$

(b)

$$\begin{aligned}Q_0 &= \sum(D_1, D_3, D_5, D_7, D_9, D_{11}, D_{13}, D_{15})\\ &= \overline{D_{14}}.\overline{D_{13}}.\overline{D_{12}}.(\overline{D_{11}}.\overline{D_{10}}.\overline{D_9}.\overline{D_8}.(\overline{D_7}.\overline{D_6}.\overline{D_5}.\overline{D_4}.\\ &\quad(\overline{D_2}.D_1 + D_3) + (\overline{D_6}.D_5 + D_7)) + \overline{D_{10}}.D_9 + D_{11}) +\\ &\quad\overline{D_{14}}.D_{13} + D_{15}\\ Q_1 &= \sum(D_2, D_3, D_6, D_7, D_{10}, D_{11}, D_{14}, D_{15})\\ &= (\overline{D_{13}}.\overline{D_{12}}.(\overline{D_{11}}.\overline{D_{10}}.\overline{D_9}.\overline{D_8}.(\overline{D_7}.\overline{D_6}.\overline{D_5}.\overline{D_4}.\\ &\quad(D_2 + D_3) + D_6 + D_7) + D_{10} + D_{11})) + D_{14} + D_{15}\\ Q_2 &= \sum(D_4, D_5, D_6, D_7, D_{12}, D_{13}, D_{14}, D_{15})\\ &= (\overline{D_{11}}.\overline{D_{10}}.\overline{D_9}.\overline{D_8}.((D_7 + D_6) + (\overline{D_7}.\overline{D_6}.(D_5 + D_4))))\\ &\quad + D_{12} + D_{13} + D_{14} + D_{15}\\ Q_3 &= \sum(D_8, D_9, D_{10}, D_{11}, D_{12}, D_{13}, D_{14}, D_{15})\\ &= D_8 + D_9 + D_{10} + D_{11} + D_{12} + D_{13} + D_{14} + D_{15}\end{aligned}$$

(c)

Fig. 4. The truth table and Boolean expression of (a) PE4, (b) PE8, and (c) PE16.

observe the complexity of expressions increases drastically as PE size varies from 4-bit to 16-bit, which possibly causes an implementation of 32-bit PE to become impracticable. Thus, only PE4, PE8, and PE16 are employed to construct large-sized PEs. Concretely, at $L$ of 64-bit, we examine ($M$, $N$) as (8, 8), (4, 16), and (16, 4).

Fig. 5(a) shows PE64 formed by two PE8s connecting in a series, namely $\mathrm{PE64}_{(8n)}$. To begin with, the input data $D$ is separated into eight 8-bit signals that are orderly put into eight 8-bit OR gates (OR8s) together with the 8-to-1 multiplexer (MUX8N). The output of MUX8N, so-called $DMUX$, is determined by $MR8$ - the position of highest priority bit of $DOR$. Following this, $MC8$, the location of the highest priority bit of $DMUX$, is obtained. The output $Q$ is derived from the bitwise OR between $MC8$ and $MR8$ that was shifted left by three bits. Additionally, if $D$ contains any 1-bit, $M$ turns into one.

Because $\mathrm{PE64}_{(8n)}$ follows the formula stated in Fig. 2, the longest delay of $\mathrm{PE64}_{(8n)}$ is approximately the sum of four individual components' delay. In fact, MUX8N has to wait until $MR8$ is ready before allocating a proper column index to $DMUX$. To reduce such delay, we employ $DOR$ as a look-ahead signal, which is illustrated in Fig. 5(b). As can be easily seen, $DOR$ cuts the longest data path, from the input of PE8_0 to the output $Q$, in two shorter paths operating in parallel. Therefore, the entire latency of $\mathrm{PE64}_{(8)}$ is likely to be fairly lowered, as compared to that of $\mathrm{PE64}_{(8n)}$. Moreover, the select signals inside MUX8 must be reassigned because of the difference in the number of bits between $MR8$ and $DOR$.

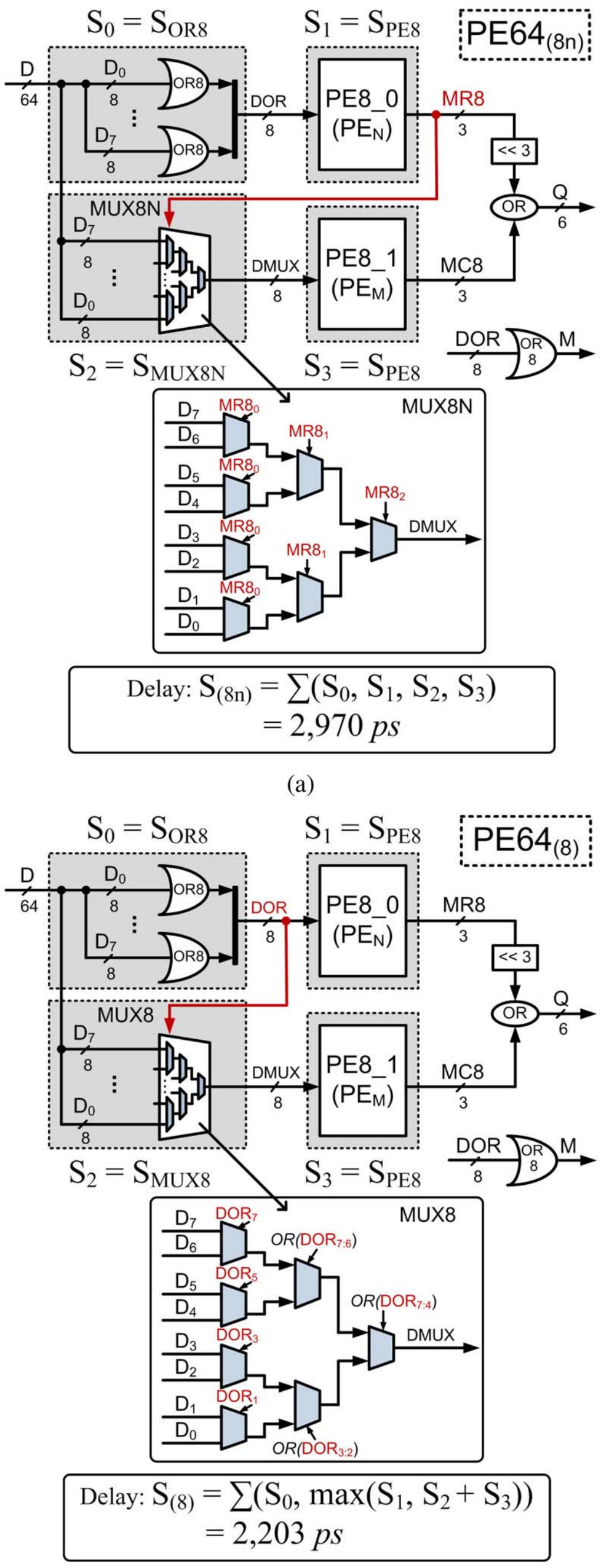

(a)

(b)

Fig. 5. The architecture of PE64 (a) without look-ahead signal and (b) with look-ahead signal.

The resource utilization of $\mathrm{PE64}_{(8)}$, hence, increases because MUX8 requires several additional OR gates.

To quickly estimate PE performance, we synthesize all OR gates, PEs, and multiplexers to observe the path delay (in terms of $ps$), from the input to the output of each circuit. The synthesis tool is configured to generate the gate-level logic under an aggressive timing constraint. Table I summarizes the synthesized results in 180-nm CMOS technology. Suppose that $S_0$, $S_1$, $S_2$, and $S_3$ are the path delays of four primary circuits in $\mathrm{PE64}_{(8n)}$ and $\mathrm{PE64}_{(8)}$. As seen in Fig. 5(a), without a look-ahead signal, the delay of $\mathrm{PE64}_{(8n)}$ is $S_{(8n)} = \Sigma(S_0, S_1, S_2, S_3) = 2{,}970$ *ps*. On the other hand, the latency PE64$_{(8)}$ is lessened as $S_{(8)} = \Sigma(S_0, \max(S_1, S_2 + S_3))$ = 2,203 *ps*. The preliminary analysis suggests that the look-ahead signal enhances the circuit performance.

TABLE I
THE NUMBER OF LOGIC STAGES

| Circuit | Delay (ps) | Circuit | Delay (ps) | Circuit | Delay (ps) |
|---|---|---|---|---|---|
| OR4 | 336 | PE4 | 680 | 16-bit MUX4 | 914 |
| OR8 | 399 | PE8 | 780 | 8-bit MUX8N | 1,011 |
| OR16 | 550 | PE16 | 980 | 8-bit MUX8 | 1,024 |
| | | | | 4-bit MUX16 | 1,070 |

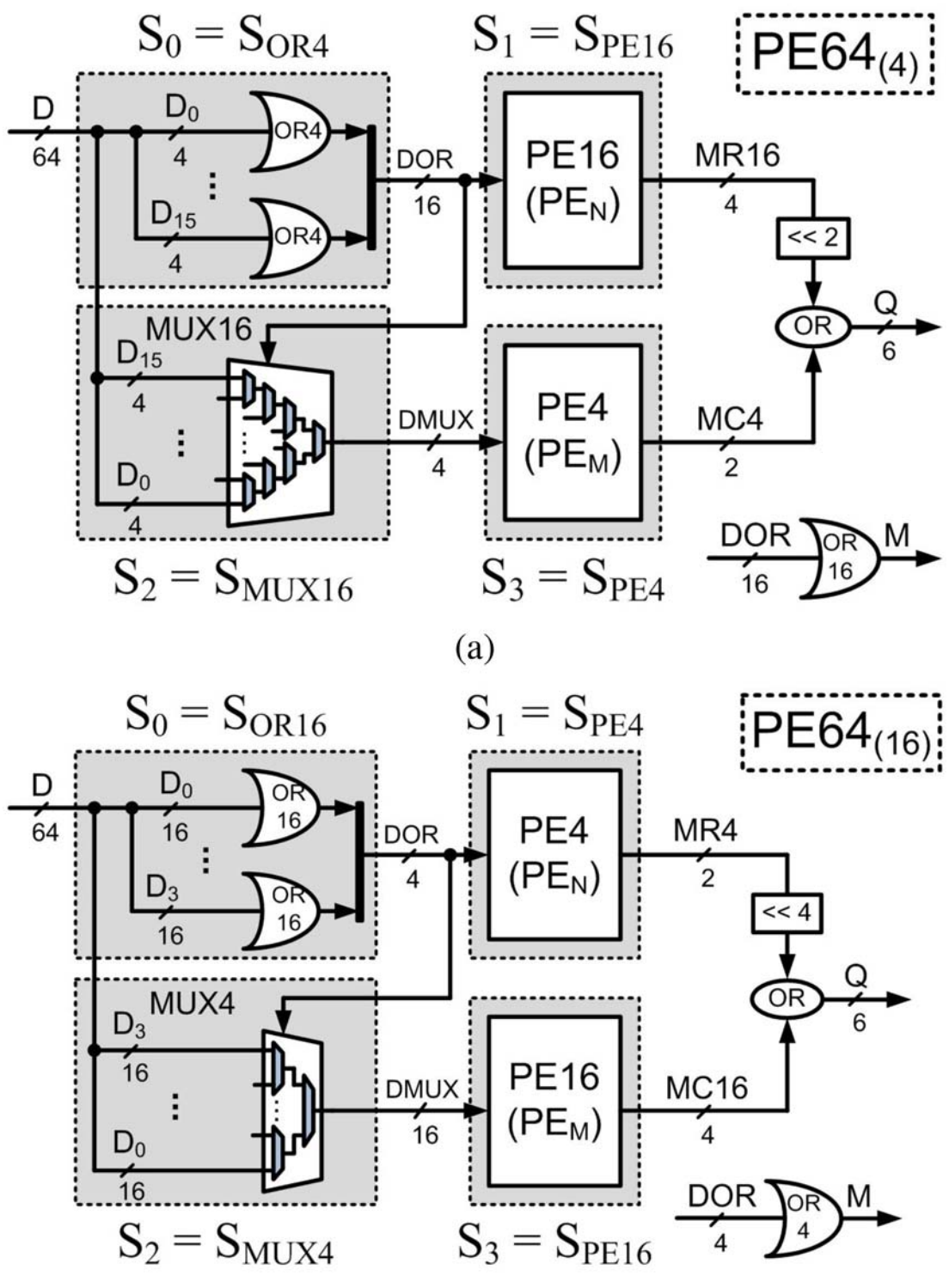

Fig. 6. The architecture of PE64 with (a) ($M$, $N$) = (4, 16) and (b) ($M$, $N$) = (16, 4).

As briefly mentioned before, in case of PE64, there are three possible pairs of ($M$, $N$), i.e., (8, 8), (16, 4), and (4, 16). The architecture of PE with ($M$, $N$) of (4, 16) and (16, 4), so-called PE64$_{(4)}$ and PE64$_{(16)}$, are defined in Fig. 6(a) and Fig. 6(b), respectively. It is noted that PE$_N$ and PE$_M$ also represent the top PE and bottom PE. In both architectures, the highest priority bit of input data $D$ is discovered in a similar vein with PE64$_{(8)}$, except the different use of OR gates, multiplexers, and the organization of PE$_N$ and PE$_M$. Using the preliminary analysis above, the path delay of PE64$_{(4)}$ is $S_{(4)}$ = 2,086 *ps*, whereas that of PE64$_{(16)}$ is $S_{(16)}$ = 2,444 *ps*. Altogether, the performance of four alternative PE64s are sorted as PE64$_{(4)}$ > PE64$_{(8)}$ > PE64$_{(16)}$ > PE64$_{(8n)}$. In other words, if PE4 is used to generate the column index ($M$ = 4), the overall performance is likely to become the best.

This preliminary analysis also implies the scalable architecture of a large-sized PE such as PE4K$_{(4)}$ that can be developed by PE4, PE16, PE64$_{(4)}$, 256-bit PE (PE256$_{(4)}$), and 1,024-bit PE (PE1K$_{(4)}$), as seen in Fig. 7. Initially, the 4,096-bit input is considered as a 1,024×4-bit array. Subsequently, PE1K$_{(4)}$ and PE4 are employed to calculate the correspondent indexes of row and column. Similarly, inside PE1K$_{(4)}$, the 1,024-bit is converted into 256×4-bit array for the next processing from PE256$_{(4)}$ and PE4. Dividing the input repeats until PE$_N$ is either PE16 or PE8. Finally, the highest priority bit is achieved from all PE outputs, based on the formula described in Fig. 2.

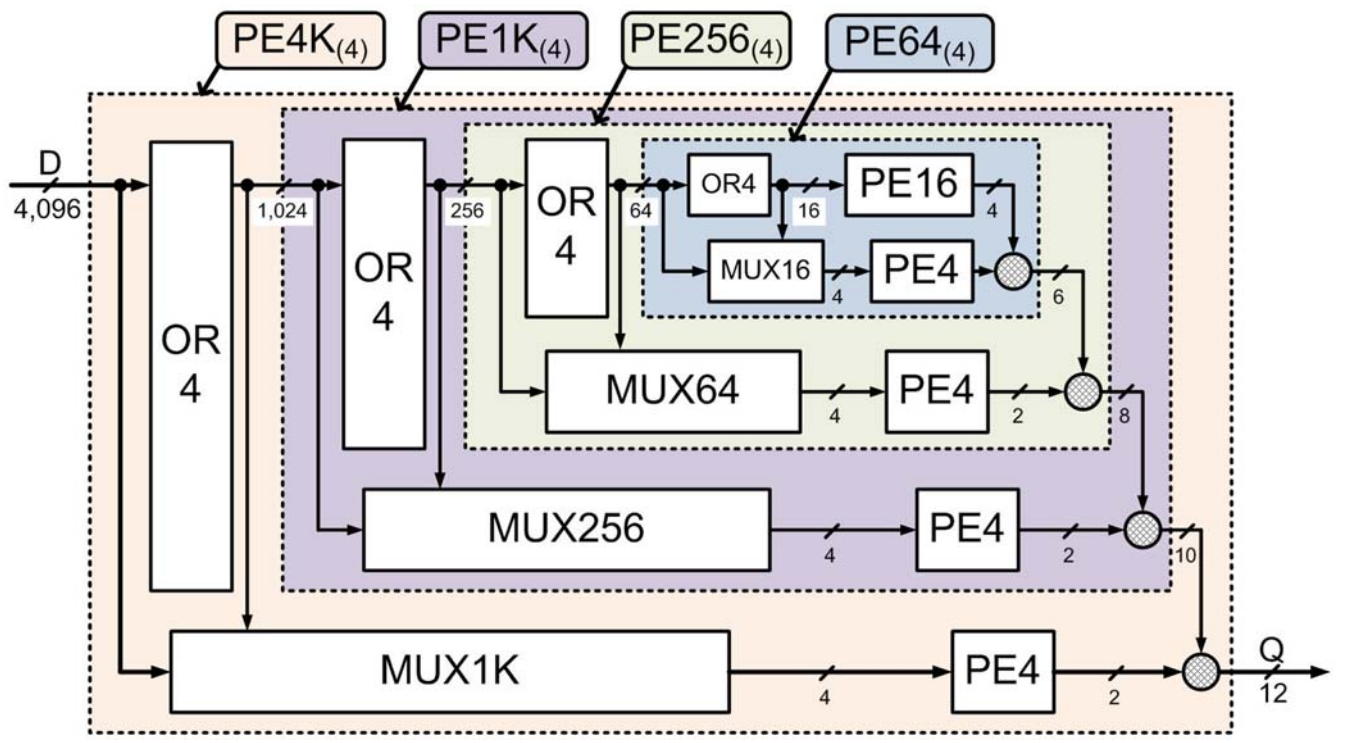

Fig. 7. The scalable architecture of PE4K$_{(4)}$.

TABLE II
THE SIMULATION RESULTS OF PROPOSED PEs

| Design | PE$_N$ / PE$_M$ | FREQ (MHz) | DEC$_L$ (%) | Transistors |
|---|---|---|---|---|
| PE4 | –/– | 1,470 | – | 48 |
| PE8 | –/– | 1,282 | -12.7 | 202 |
| PE16$_{(4)}$ | PE4 / PE4 | 909 | – | 406 |
| PE16 | –/– | 1,020 | -20.4 | 396 |
| PE32$_{(8)}$ | PE4 / PE8 | 645 | – | 626 |
| PE32$_{(4)}$ | PE8 / PE4 | 757 | -25.2 | 1,046 |
| PE64$_{(16)}$ | PE4 / PE16 | 520 | – | 1,702 |
| PE64$_{(8n)}$ | PE8 / PE8 | 526 | – | 2,128 |
| PE64$_{(8)}$ | PE8 / PE8 | 555 | – | 2,464 |
| PE64$_{(4)}$ | PE16 / PE4 | 649 | -14.2 | 1,768 |
| PE128$_{(16)}$ | PE8 / PE16 | 480 | – | 3,804 |
| PE128$_{(8)}$ | PE16 / PE8 | 510 | – | 3,536 |
| PE128$_{(4)}$ | PE32$_{(4)}$ / PE4 | 595 | -8.3 | 2,292 |
| PE256$_{(4)}$ | PE64$_{(4)}$ / PE4 | 520 | -12.6 | 9,448 |
| PE512$_{(4)}$ | PE128$_{(4)}$ / PE4 | 462 | -11.1 | 13,256 |
| PE1K$_{(4)}$ | PE256$_{(4)}$ / PE4 | 434 | -6.1 | 18,960 |
| PE2K$_{(4)}$ | PE512$_{(4)}$ / PE4 | 416 | -4.1 | 42,722 |
| PE4K$_{(4)}$ | PE1K$_{(4)}$ / PE4 | 370 | -11.1 | 70,502 |

## IV. PERFORMANCE ANALYSIS

Various PEs whose sizes vary from 4-bit to 4,096-bit are implemented in 180-nm CMOS technology. Their performance is evaluated by both FREQ and resource utilization, which are obtained from the post-place-and-route simulation results at 1.8 V. The simulation values in Table II point out three main findings:

- Firstly, 1D-to-2D conversion usage evidently improves the deterioration of performance at large PE sizes. In fact, assume DEC$_L$ is the percentage decrease of FREQ between PE$_L$ and PE$_{L/2}$, it is easy to see the major difference between DEC$_8$ and DEC$_{16}$, whose circuits are directly built from the truth tables. On the contrary, from DEC$_{64_{(4)}}$ to DEC$_{4K_{(4)}}$, the mean value is approximately 11% whenever PE size is doubled.

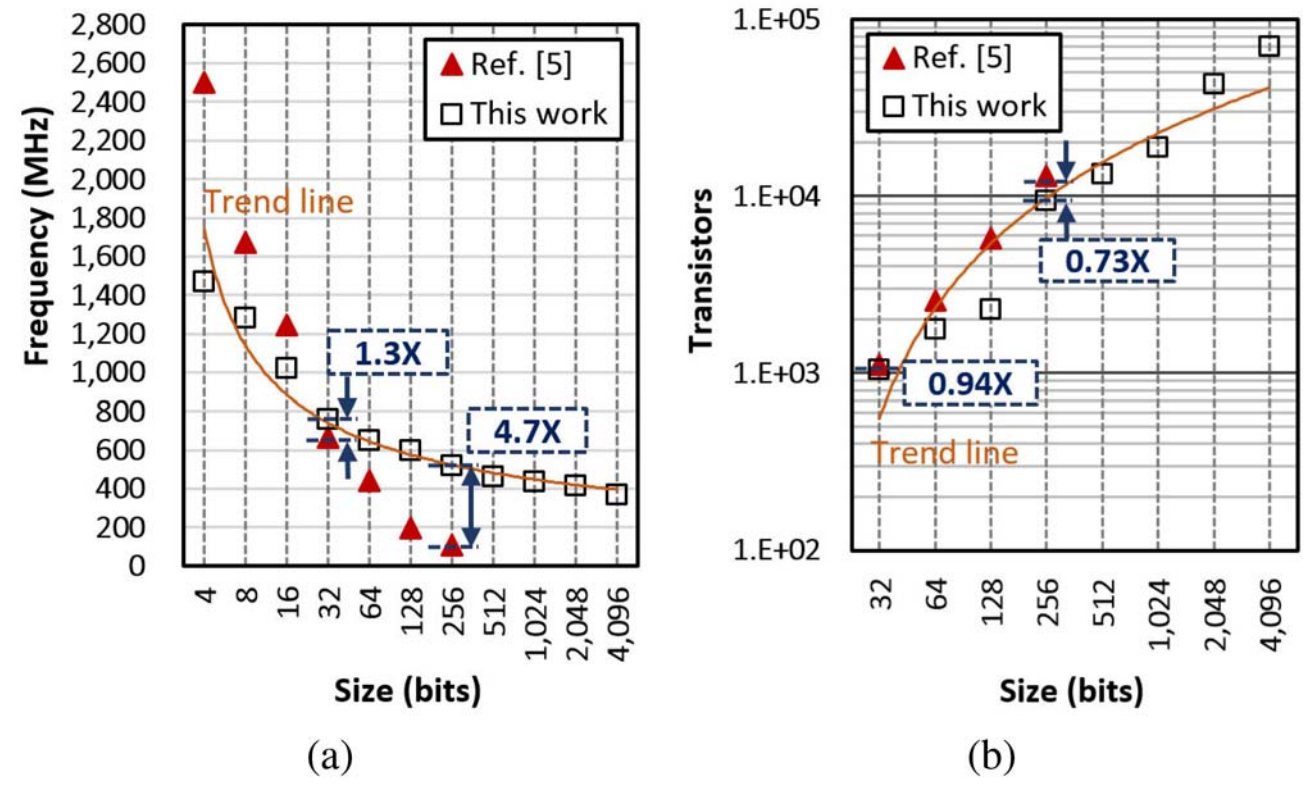

Fig. 8. The comparison with [5].

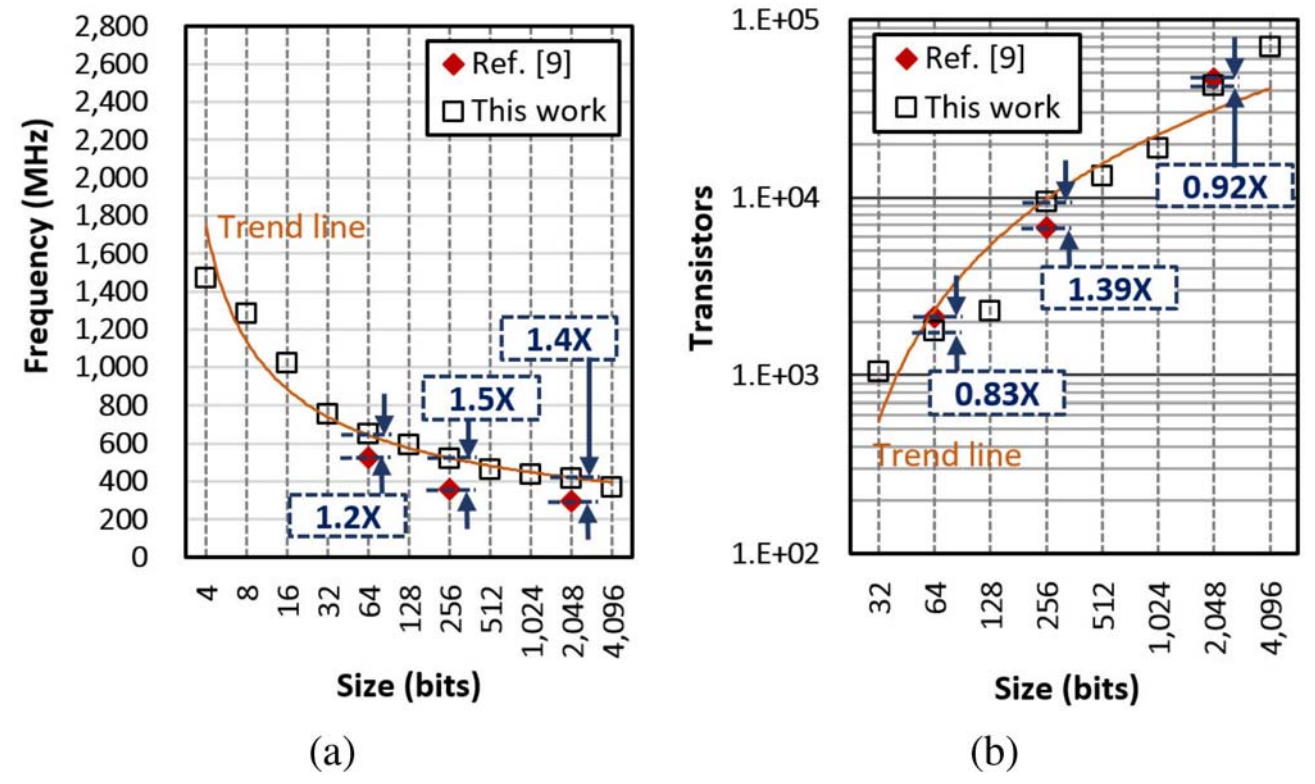

Fig. 9. The comparison with [9].

• Secondly, the look-ahead signal usage fairly contributes to FREQ enhancement. Taking an example of $PE64_{(8n)}$ and $PE64_{(8)}$, the FREQ of the latter increases approximately 5.2%. The improvement is not as high as the preliminary analysis because in the real implementation, we applied flat-design synthesis in each PE. In this mode, hierarchical boundaries are removed, thereby reducing the levels of logic and improving the timing of each PE.

• Thirdly, the organization of $PE_N$ and $PE_M$ clearly affects the outcome of a large-sized PE. For example, $PE64_{(4)}$ achieves the highest FREQ while $PE64_{(16)}$ obtains the lowest FREQ, which is identical to the above preliminary analysis. Hence, only large-sized PEs with $M = 4$ are compared with other previous works.

In comparison with [5], which was simulated in 150-nm CMOS technology, current designs gradually become better when PE sizes vary from 32-bit to 256-bit. As seen in Fig. 8(a), FREQ of $PE32_{(4)}$ is only 1.3 times as high as that of [5], whereas at PE size of 256-bit, the difference of FREQ remarkably increases to 4.7 times. Moreover, according to Fig. 8(b), the transistor count of $PE32_{(4)}$ and $PE256_{(4)}$ are only 0.94 times and 0.73 times, as compared to those in [5].

In addition, Fig. 9 depicts the comparison of FREQ and transistor count between two works in 180-nm CMOS technology when PE sizes vary from 64-bit to 2,048-bit. Because the architecture of PE4, PE8, and PE16 are identical in both works, their FREQs and transistor count are unchanged. In [9], PE64 shares the same architecture with $PE64_{(8n)}$, where $(M, N) = (8, 8)$ is a non-optimal configuration and look-ahead signal is unused. In a similar vein, PE256 is constructed by two PE16s. PE2K, however, is formed by 32 PE64s operating in parallel together with one central PE32. In fact, its architecture is similar to the PE4K's, as demonstrated in Fig. 3(b).

As seen in Fig. 9(a), the FREQs of $PE64_{(4)}$, $PE256_{(4)}$, and $PE2K_{(4)}$ are 1.2 times, 1.5 times, and 1.4 times as high as those in [9]. When it comes to logic utilization, $PE64_{(4)}$ and $PE2K_{(4)}$ cost fewer transistors than PE64 and PE2K, respectively, as depicted in Fig. 9(b). However, the power consumption of $PE64_{(4)}$ and $PE2K_{(4)}$ are 20.6% and 29.9% as high as those in [9]. Nevertheless, the resource and power consumption will be considered as the future work as this brief mainly concentrates on the high-performance architecture. In short, our architecture offers higher performance as compared to [5] and [9].

## V. Conclusion

We have presented a method to develop a scalable architecture of high-performance large-sized PEs. By employing 1D-to-2D conversion, the deterioration of performance at large PE sizes is improved significantly, i.e., FREQ reduces gradually 11% whenever PE size is doubled. Further, at $PE64_{(4)}$, $PE256_{(4)}$, and $PE2K_{(4)}$, our FREQs are 1.2 times, 1.5 times, and 1.4 times as high as those of prior work.

## References

[1] C.-H. Huang, J.-S. Wang, and Y.-C. Huang, "Design of high-performance CMOS priority encoders and incrementer/decrementers using multilevel lookahead and multilevel folding techniques," *IEEE J. Solid-State Circuits*, vol. 37, no. 1, pp. 63–76, Jan. 2002.

[2] S.-W. Huang and Y.-J. Chang, "A full parallel priority encoder design used in comparator," in *Proc. IEEE Int. Midwest Symp. Circuits Syst.*, Seattle, WA, USA, 2010, pp. 877–880.

[3] C. Kun, S. Quan, and A. Mason, "A power-optimized 64-bit priority encoder utilizing parallel priority look-ahead," in *Proc. IEEE Int. Symp. Circuits Syst.*, vol. 2. Vancouver, BC, Canada, 2004, pp. II-753–II-756.

[4] M. Faezipour and M. Nourani, "Wire-speed TCAM-based architectures for multimatch packet classification," *IEEE Trans. Comput.*, vol. 58, no. 1, pp. 5–17, Jan. 2009.

[5] S. Abdel-Hafeez and S. Harb, "A VLSI high-performance priority encoder using standard CMOS library," *IEEE Trans. Circuits Syst. II, Exp. Briefs*, vol. 53, no. 8, pp. 597–601, Aug. 2006.

[6] D. Balobas and N. Konofaos, "Low-power, high-performance 64-bit CMOS priority encoder using static-dynamic parallel architecture," in *Proc. IEEE Int. Conf. Modern Circuits Syst. Technol. (MOCAST)*, Thessaloniki, Greece, 2016, pp. 1–4.

[7] D.-H. Le, K. Inoue, M. Sowa, and C.-K. Pham, "An FPGA-based information detection hardware system employing multi-match content addressable memory," *IEICE Trans. Fundam. Electron. Commun. Comput. Sci.*, vol. E95.A, no. 10, pp. 1708–1717, Oct. 2012.

[8] S. K. Maurya and L. T. Clark, "A dynamic longest prefix matching content addressable memory for IP routing," *IEEE Trans. Very Large Scale Integr. (VLSI) Syst.*, vol. 19, no. 6, pp. 963–972, Jun. 2011.

[9] X.-T. Nguyen, H.-T. Nguyen, and C.-K. Pham, "An FPGA approach for high-performance multi-match priority encoder," *IEICE Electron. Exp.*, vol. 13, no. 13, pp. 1–9, Jun. 2016.